\documentclass[aps,pra,twocolumn,groupedaddress,byrevtex]{revtex4}
\usepackage{graphicx}
\newcommand{\im}{\mathrm{ i}}
\newcommand{\xv}{\vec{x}}
\newcommand{\q}{\vec{q}}
\newcommand{\xp}{\vec{x}'}

\begin{document}
\title{Coherent imaging with pseudo-thermal incoherent light}
\author{A. Gatti, M. Bache, D. Magatti, E. Brambilla, F. Ferri, and
L.A. Lugiato}
\affiliation{INFM, Dipartimento di Fisica e Matematica, Universit{\`a}
 dell'Insubria, Via Valleggio 11, 22100 Como, Italy}
\date{February 15, 2005} 
%

%\pacs{42.50.Dv, 42.50-p, 42.50.Ar}
\begin{abstract}
  We investigate experimentally fundamental properties of coherent
  ghost imaging using spatially incoherent beams generated from a
  pseudo-thermal source. A complementarity between the coherence of
  the beams and the correlation between them is demonstrated by
  showing a complementarity between ghost diffraction and ordinary
  diffraction patterns. In order for the ghost imaging scheme to work
  it is therefore crucial to have incoherent beams. The visibility of
  the information is shown for the ghost image to become better as the
  object size relative to the speckle size is decreased, and therefore
  a remarkable tradeoff between resolution and visibility exists.
  %Instead, the visibility of the diffraction pattern is shown to
  %become better as the absolute object size is increased. 
  The experimental conclusions are backed up by both theory and numerical
  simulations.
\end{abstract}
\maketitle
\section{Historical overview and introduction}
A decade has passed since the first experimental observation
of unusual interference fringes in the coincidence counts of photon
pairs \cite{strekalov:1995,ribeiro:1994}. Signal and idler photons
produced by parametric down-conversion (PDC) were spatially separated
and in the signal photon arm a double slit was inserted. While no
first order interference pattern was visible behind the slit, an
interference pattern was observed in the coincidence count { by
  scanning the idler photon detector position}. To this phenomenon it
was given the name {\em ghost diffraction}. Shortly after a {\em ghost
  image} experiment was performed \cite{pittman:1995}, showing a sharp
image of an object placed in the signal arm by registering the
coincidence counts as a function of the idler photon position.
\par 
In the interpretations of the experiments, the quantum nature of the
source there employed was emphasized, although the Authors of
\cite{pittman:1995} suggested that {\em "it is possible to imagine
  some type of classical source that could partially emulate this
  behavior"}.  Several years passed before a systematic theory of
ghost imaging (GI) started to be developed, and soon a lively debate
arose discussing the role of entanglement versus classical correlation
in GI schemes. In the first theoretical papers by the Boston group
\cite{abouraddy:prl-josab-osa}, it was claimed that entanglement was a
crucial prerequisite for achieving GI, and in particular coherent GI:
{\em "the distributed quantum-imaging scheme truly requires
  entanglement in the source and cannot be achieved using a classical
  source with correlations but without entanglement." } Soon after, at
Rochester University a ghost image experiment was performed exploiting
the classical angular correlation of narrow laser pulses
\cite{bennink:2002a}.  This fueled the debate: which are the features
of ghost imaging that truly requires entanglement? The debate was
continued by the paper \cite{gatti:2003}, where some of us showed that
a classical GI scheme can indeed produce either the object image or
the object diffraction pattern, but suggested that both cannot be
produced without making changes to the source or the object arm setup.
We argued that only entangled beams can give both results by only
changing the setup in the reference arm (the one where the object is
not present). While by now we know this is not true, at that time it
was in partial agreement with \cite{abouraddy:prl-josab-osa} and
\cite{bennink:2002a}.  When the Rochester group recently completed the
results by showing that also the object diffraction pattern can be
reconstructed using classically correlated beams \cite{bennink:2004},
they had indeed to change the setup (the object location, the lens
setup as well as the detection protocol).

Our claim in \cite{gatti:2003} originated from the fact that only
entangled beams can have simultaneously perfect spatial correlation in
both the near and the far field (in both position and momentum of the
photons), and no classical beams can mimic this \cite{brambilla:2004}.
In the same spirit, recent experimental works
\cite{howell:2004,bennink:2004,dangelo:2004}, brilliantly pointed out
a momentum-position realization of the Einstein-Podolsky-Rosen (EPR)
paradox using entangled photon pairs produced by PDC. The product of
conditional variances in momentum and position was there shown to be
below the EPR bound that limits the correlation of any classical
(non-entangled) light beam. Based on these results, the Authors of
\cite{bennink:2004} proposed the same EPR bound as a limit for the
product of the resolutions of the images formed in the near and in the
far-field of a given classical source, and both
\cite{bennink:2004,dangelo:2004} argued that in ghost imaging schemes
entangled photons allow to achieve a better spatial resolution than
any classically correlated beams.

\par
This was the state of the art, when  some of us had an idea leading to
a ghost imaging protocol with classical thermal-like beams. Inspired
by the fact that the marginal statistics of the signal or idler beam
from PDC is of thermal nature, we asked ourselves what would have been
the result of splitting a thermal beam on a macroscopic beam splitter
and using the two outgoing beams for ghost imaging. Honestly speaking, we expected at that time that this would have lead to identify relevant differences with quantum entangled beams, where the correlation is of microscopic origin. The picture that came out was however rather different. The two output beams of the
beam splitter are obviously each a true copy (on a classical level) of the input beam:  if there is a speckle
at some position in the input beam, then each output beam has also a speckle at the same position.  Hence the beam
splitter has created beams with a strong spatial correlation between them, while each beam on its own
is spatially incoherent. In theoretical works \cite{thermal-oe} we showed that this correlation is preserved upon propagation (so it is present
both in the near and the far-field planes), and that the beams  could therefore be used to perform GI exactly in the same way as the entangled beams from PDC. Actually a very close formal analogy was demonstrated between GI with thermal and PDC beams, which implied that classically  correlated beams were able to emulate {\em all} the relevant features of quantum GI, with the only exception of the visibility \cite{thermal-oe}.

\par
Thus, we actually had to
conclude that what we had written in \cite{gatti:2003} was not wrong but not very correct either. What we failed to recognize there is that ghost imaging protocols do not need a perfect correlation at all: with the imperfect (shot-noise limited) spatial correlation of thermal beams both the object image and the object diffraction pattern can be reproduced without
making any changes to the source and only changing the reference arm. Moreover the formal analogy between thermal and PDC beams suggested that identical performances with respect to spatial resolution should be achieved by the quantum and classical protocols, provided that the spatial coherence properties of the two sources were similar. This
was obviously a controversial result compared to what was published until that time \cite{abouraddy:prl-josab-osa, bennink:2002a,gatti:2003} and in order to be accepted needed an experimental confirmation. We recently provided this \cite{ferriexp-osa} showing  high-resolution ghost image and ghost diffraction experiments, performed by using a single source of pseudo-thermal speckle light
divided by a beam splitter. As predicted, it was possible to pass from the image by only  changing the optical setup in the reference arm, while leaving the object arm untouched. Moreover, the product of spatial resolutions of the ghost image and ghost diffraction experiments was there shown to overcome the EPR bound  which was proposed to be achievable only with entangled photons by former literature \cite{bennink:2004,dangelo:2004}. The origin of the apparent contradiction with the former literature was there identified, by recognizing that the spatial resolution of GI protocols do not coincide in general with the conditional variance, so that the product of the near and far-field resolution is free from any EPR separability bound.

\par
The idea of using pseudo-thermal light for GI had some enthusiastic followers 
\cite{cai:2004d,cao:2004} with proposals for X-ray diffraction
\cite{cheng:2004}, some partially converted fans, with experiments characterizing a pseudo-thermal source of photon pairs\cite{scarcelli:2004}
and using them for realizing a ghost image \cite{valencia:2004}, and some sceptics \cite{abouraddy:2004}.
The use of pseudo-thermal light in GI schemes inspired also a topic which became of a certain interest, known as  "quantum lithography with classical beams", or "sub-wavelength interference with classical beams". The quantum version of this  started with the famous paper by Boto et al. \cite{boto:2000} claiming that N-photon entangled states
could be used for improving the resolution of lithography by a factor of N. A proof-of-principle experiment using N = 2 in the PDC case was provided by \cite{dangelo:2001} where a halving  of the period of the interference fringes was observed in a  "ghost 
diffraction" pattern.  In \cite{gatti:2003} some of us observed  that the same effect may be observed when thermal-like beams
are used, and that in both the entangled and thermal case the sub-wavelength interference relies on a
simple geometrical artifact. We therefore questioned whether the Shih experiment really proves Boto's entangled protocol. 
Sub-wavelength interference using thermal beams was then theoretically discussed in \cite{wang:2004}, and experimentally demonstrated \cite{scarcelli:2004a}.

\par
%\section{Introduction}
In this paper we continue the experimental investigations started
in \cite{ferriexp-osa}. The main result  there established was that high 
resolution ghost image and ghost diffraction could  be achieved with the same classical source, with the product of resolutions well behind the EPR bound proposed by \cite{bennink:2004}. Here, we shall
investigate first of all a fundamental complementarity between
coherence and correlation which exists for ghost imaging schemes. 
Only when the beams are spatially incoherent the correlation functions allows to retrieve information about the object (ghost image or ghost diffraction), while the information disappears as the spatial coherence of the beams increases. This is just the opposite of what happens for direct detection of the light behind the object, where fully coherent information can be obtained only for spatially coherent beams. Thus,
the spatial incoherence plays an essential role for realizing ghost
imaging, while instead the Hanbury-Brown--Twiss interferometer
\cite{hanburybrown:1956} for determining the stellar diameter relies
on coherence gained by propagation. Secondly, we will investigate
visibility and signal-to-noise ratio in ghost imaging with thermal
light, and highlight a tradeoff between visibility and resolution when
reconstructing the information.

The paper is organized as follows: In Sec.~\ref{sec:Descr-exper} the
experiment is described, while Sec.~\ref{sec:formal} introduces the
formalism and review the formal analogy between classical and
quantum ghost imaging. In addition, the relation between visibility and
signal-to-noise ratio is discussed. In Sec.~\ref{sec:Spat-coher-prop}
the spatial coherence properties of the beams are investigated and
experimentally characterized. Section~\ref{sec:diff} focusses on the
ghost diffraction setup and shows the complementarity between coherence
and correlation.  Section~\ref{sec:image} focusses on the ghost
image, and discusses visibility. In Sec.~\ref{sec:Numerical-results}
numerical results are presented which provide a more detailed insight
into the results of the experiment. Finally, the conclusions are drawn
in Sec.~\ref{sec:Conclusions}.

%%%%%%%%%%%%%%%%%%%%%%%
\section{Description of the experiment}
\label{sec:Descr-exper}
\begin{figure}[ht] 
\centerline{    
    \scalebox{.5}{\includegraphics*{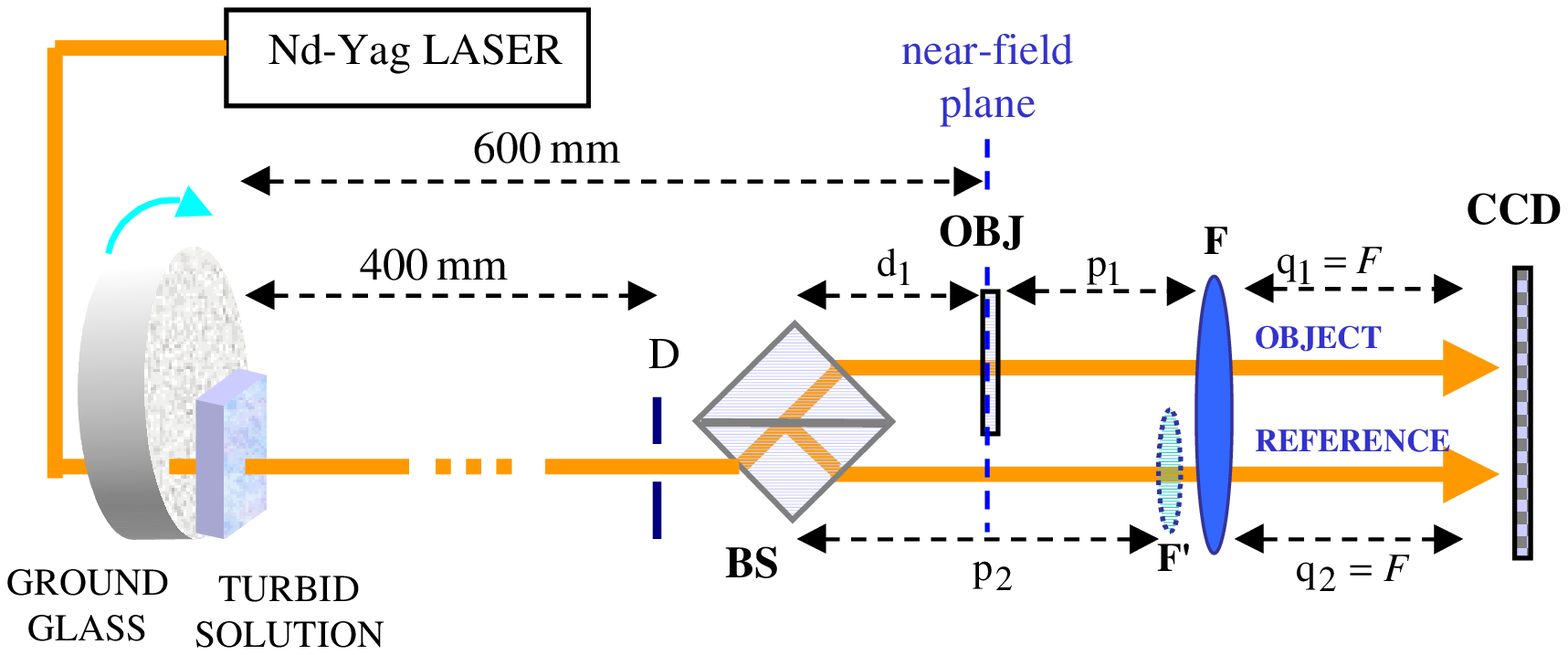}} }
\caption{Scheme of the setup of the experiment (see text for details). }
\label{fig:setup}
\end{figure}

The experimental setup is similar to that of Ref.\cite{ferriexp-osa} and
is sketched in Fig.\ref{fig:setup}.  The source of pseudo-thermal
light is provided by a scattering medium illuminated by a laser beam.
The medium is a slowly rotating ground glass placed in front of a
scattering cell containing a turbid solution of $3\,\mu$m latex
spheres.  When this is illuminated with a large collimated Nd-Yag
laser beam ($\lambda = 0.532\, \mu$m, diameter $D_0 \approx 10$ mm),
the stochastic interference of the waves emerging from the source
produces at large distance ($z \approx 600$ mm) a time-dependent
speckle pattern, characterized by a chaotic statistics and by a
correlation time $\tau_{coh}$ on the order of 100 ms (for an
introduction to laser speckle statistics see e.g.\cite{goodman:1975}).
Notice that the ground glass can be used alone to produce chaotic
speckles, whose correlation time depends on the speed of rotation of
the ground-glass disk and on the laser diameter, as in classical
experiments with pseudo-thermal light \cite{Martienssen, Arecchi}.
Indeed, in some part of the experiments described in the following it
will be used alone. This however presents the problem that the
generated speckle patterns reproduce themselves after a whole tour of
the disk, which can be partially avoided by shifting laterally the
disk at each tour. The turbid solution provides an easy way to
generate a truly random statistics of light, because of the random
motion of particles in the solution, allowing a huge number of
independent patterns to be generated and used for statistics. Notice
that the turbid medium cannot be used alone, because a portion of the
laser light would not be scattered, thus leaving a residual coherent
contribution.

At a distance $z_0=400$ mm from the thermal source, a diaphragm of diameter $D=3$ mm selects an angular portion of the speckle pattern,
allowing the formation of an 
almost collimated speckle beam  characterized by a huge number (on the order of $10^4$) of speckles of size $\Delta x \approx \lambda z_0/D_0 \approx 21 \,\mu$m \cite{goodman:1975}. The speckle beam is separated by the beam splitter (BS) into two
"twin" speckle beams, that exhibit a high  (although classical)  level of spatial correlation \cite{thermal-oe}. 
The two beams emerging from the BS have slightly non-collinear propagation directions, and  illuminate  two different non-overlapping portions of the charged-coupled-device (CCD) camera. The data are 
acquired with an exposure time ($1-3$ ms) much shorter than $\tau_{coh}$, allowing the recording of high contrast speckle patterns. The frames are 
taken at a rate of 1-10 Hz, so that each data acquisition corresponds to 
uncorrelated speckle patterns. 

In one of the two arms (the object arm 1) an object about which we
need to extract information is placed. The object plane, located at a
distance $200$ mm from the diaphragm, will be taken  as the reference
plane, and referred to as  the {\em near-field} plane (this is not to
be confused with the source near-field,  as the object plane is in the
far zone with respect to the source). The optical setup of the object
arm is kept fixed, and consists of a a single lens of focal length
$F=80\,$mm,  placed  at a distance $p_1$ after the object and $q_1=F$
from the CCD. In this way the CCD images the far-field plane with
respect to the object.

We shall consider two different setups for the reference arm 2. 
In the {\em ghost-diffraction} setup, the reference beam passes 
through the same lens $F$ as the object beam, located at a distance $q_2=F$ from the CCD. In Ref.\cite{ferriexp-osa}, the spatial cross-correlation  function of  the reference and object arm 
intensity distributions was calculated, 
 and showed a sharp reproduction of the diffraction pattern of the
 object, comparable with the diffraction pattern obtained  by
 illuminating the object with the laser light (see
 Sec. \ref{sec:diff}).

In the {\em ghost-image} setup, without changing anything in the object arm,  an additional lens of focal length $F^{\prime}$ is  inserted
in the reference arm  immediately before $F$. The total focal length $F_2$ of the two-lens system is smaller
than its distance from the CCD, being $ \frac{1}{F_2}\approx\frac{1}{F}+ \frac{1}{F^{\prime}}$. It was thus possible to locate the position of the plane conjugate to
the detection plane, by temporarily inserting the object in the reference arm 
and determining the position that produced a well focussed image on the CCD with laser illumination. The object was then translated in the object arm. 
The distances in the reference arm approximately obey a thin lens equation of the form 
${1}/{(p_2-d_1)} + {1}/{q_2} \approx {1}/{F_2}$ 
\footnote{This is only approximately true because the two-lens system is equivalent to a thick lens rather than a thin lens.}, providing a magnification factor $m=1.2$.
In Ref.\cite{ferriexp-osa} the intensity distribution of the reference arm was correlated with the total photon counts of the object arm 
showing in this case  a well-resolved reproduction of  the image of the object (see also Sec. \ref{sec:image}). 
Thus, the setups allows  a high-resolution reconstruction of both the image and the diffraction pattern of the object by using a single source of classical light. The passage from the diffraction pattern to the image is performed  by only operating on the optical setup of the reference arm, which gives evidence for the  the distributed character of the correlated imaging with thermal light.
%%%%%%%%%%%%%%%%%%%%%%%%%%%%%%%%%%%%%%%%%%%%
\section{Formal equivalence of ghost imaging with thermal beams and the two-photon entangled source} \label{sec:formal}
The basic theory behind the setup shown in Fig.~\ref{fig:setup} has
been explained in detail in Ref.~\cite{thermal-oe}. The
starting point is the input-output relations of the beam splitter 
\begin{equation}
b_1 (\xv )= t a(\xv) + r v (\xv) \, , \quad 
b_2 (\xv )= r a (\xv)+ t v(\xv) 
\label{BS} 
\end{equation}
where $b_1$ and $b_2$ are the object and reference beams emerging from the BS, $t$ and $r$ are the transmission and reflection coefficients of
the BS, $a$ is the boson operator of the speckle field and $v$ is a vacuum field
uncorrelated from $a$. The state of $a$ is a thermal mixture,
characterized by a Gaussian field statistics, in which any correlation
function of arbitrary order is expressed via the second order
correlation function
\begin{equation}
\Gamma(\xv,\xp)=
\langle a^\dagger (\xv) a (\xp) \rangle
\label{gamma}
\end{equation}
Since the collection time of our measuring apparatus is much smaller
than the time $\tau_{coh}$ over which the speckle fluctuate, all the
beam operators are taken at equal times, and the time argument is
omitted in the treatment.  Notice  that we are dealing with classical
fields, so that the field operator $a$ could be  replaced by a
stochastic c-number field, and the quantum averages by statistical
averages over independent data acquisitions. However, we prefer to
keep a quantum formalism in order to outline the parallelism with the
quantum entangled beams from PDC.

The fields at the detection planes are given by $c_i (\xv_i) = \int
{\rm d} \xv' h_i( \xv_i,\xv') b_i (\xv') + L_i (\xv_i)$, where $L_i$
account for possible losses in the imaging systems, and $h_1 $, $h_2$
are the impulse response function describing the optical setups in the two arms. 
The object
information is extracted by measuring the spatial correlation function
of the intensities $\langle I_1 (\xv_1) I_2 (\xv_2) \rangle$,
where 
$ I_i (\xv_i)$ are operators associated to the number of photo counts 
over the CCD pixel located at  $\xv_i$ in the i-th beam. All the information about the object is contained in the correlation function of intensity fluctuations, which is calculated   
by subtracting the {\em background} term $\langle I_1 (\xv_1)\rangle \langle I_2 (\xv_2) \rangle \  $: 
\begin{equation}
G(\xv_1, \xv_2) = \langle I_1 (\xv_1) I_2 (\xv_2) \rangle - \langle I_1 (\xv_1)\rangle \langle I_2 (\xv_2) \rangle \; .
\label{eq6}
\end{equation}
The main result obtained in \cite{thermal-oe} was
\begin{eqnarray}
G(\xv_1, \xv_2) = |rt|^2
\nonumber\\\times
\left| \int {\rm d} \xp_1
\int {\rm d} \xp_2  h_1^* (\xv_1, \xp_1) h_2 (\xv_2, \xp_2) \Gamma(\xv,\xp)
\right|^2 \; ,
\label{eq12}
\end{eqnarray}
Equation (\ref{eq12}) has to be compared with the analogous result obtained for PDC beams \cite{gatti:2003}:
\begin{eqnarray}
  G_{\rm pdc}(\xv_1, \xv_2)  = 
\nonumber\\\times
\left| \int d \xp_1
\int d \xp_2  h_1 (\xv_1, \xp_1) h_2 (\xv_2, \xp_2) \Gamma_{\rm
  pdc}(\xp_1,\xp_2) 
\right|^2,
\label{eq:pdc}
\end{eqnarray}
where $1$ and $2$ label the signal and idler down-converted fields $a_1$, $a_2$, and 
\begin{equation}
\Gamma_{\rm
  pdc}(\xp_1,\xp_2) = \langle a_1 (\xp_1) a_2 (\xp_2) \rangle 
\label{gamma:pdc}
\end{equation}
is the second order field correlation between the signal and idler (also called biphoton amplitude).
As already outlined in \cite{thermal-oe}, ghost imaging with correlated thermal beams, described by Eq.(\ref{eq12})  presents a deep analogy (rather than a duality) with ghost imaging with entangled beams, described by Eq.(\ref{eq:pdc}):
(a) both are \textit{coherent imaging systems}, which is crucial for
observing interference from an object, and in particular interference from a  phase object; (b) both
perform similarly if the beams have similar spatial coherence properties, that is if $\Gamma$ and $\Gamma_{\rm pdc}$ have
similar properties. They differ in a) the presence of $h_1^*$ at the place of $h_1$, which implies some non fundamental geometrical differences in the setups to be used and b) the visibility, which can be close to unity only in the in the coincidence count regime of PDC.
We define the visibility of the information as
\begin{equation}
{\cal V}= 
%\frac{  \left. G (\xv_1, \xv_2) \right|_{max}   } 
%{ \left. \left[ \langle I_1(\xv_1) \rangle \langle I_2(\xv_2)  \rangle +  
% G(\xv_1, \xv_2)\right] \,  \right|_{max}}  \, . 
 \frac{  G_{max}      } 
{\langle I_1I_2  \rangle_{max}     }   =
 \frac{  G_{max}    } 
{ \left[ \langle I_1 \rangle \langle I_2 \rangle +  G \right]_{max}    }     \, . 
\label{visibility}
\end{equation} 
In the thermal case $ G(\xv_1, \xv_2) \le \langle I_1(\xv_1) \rangle \langle I_2( \xv_2)  \rangle $ so that the visibility is never above $\frac{1}{2}$. Conversely, in the PDC case it is not difficult to verify 
that the ratio $G_{\rm pdc}/ \langle I_1 \rangle \langle I_2   \rangle $ scales as $ 1 + \frac{1}{\langle n \rangle}$, where $\langle n \rangle $ is the mean photon number per mode (see e.g. Ref. \cite{thermal-oe}b). Only in the coincidence-count regime, where $ \langle n \rangle \ll 1  $, the visibility can be close to unity, while bright entangled beams with $\langle n \rangle \gg 1$  show a similar visibility as the classical beams. However, despite never being above $\frac{1}{2}$
in the classical case, we have shown \cite{thermal-oe,ferriexp-osa} that the visibility is sufficient to 
efficiently retrieve information. 

The visibility is an important parameter in determining  the signal-to-noise ratio (SNR) associated to a ghost imaging scheme (see also \cite{cheng:2004a}). Intuitively, one expects that the noise associated to a measurement of $ I_1 I_2  $ is proportional to 
$\langle I_1 I_2\rangle$, being the statistics of thermal nature. This noise also  affects the retrieval of the ghost image or of the ghost diffraction in a single measurement, because this  is obtained from  $I_1 I_2 $ by subtracting the background term. Hence $SNR \propto G/\langle I_1 I_2\rangle$, and the visibility defined by Eq.(\ref{visibility}) gives an estimate of the signal-to-noise ratio  of a ghost imaging scheme. 
This picture is confirmed by a  more quantitative calculation, not reported here, performed by using the standard properties of Gaussian statistics. By defining $\Delta G = \sqrt{\langle O^2\rangle - \langle O\rangle^2 }$, with $O=I_1 I_2 -\langle I_1\rangle \langle I_2 \rangle$, where $G := G(\xv_1,\xv_2)$, $I_i := I_i (\xv_i)$, we obtained  $\Delta G  \approx \sqrt{3 \langle I_1  I_2 \rangle ^2+ 8 G \langle I_1 \rangle \langle I_2\rangle} $, where quantum corrections have been neglected. If the visibility is small, as it is often the case, this reduces to 
$\Delta G  \approx \sqrt{3} \langle I_1  I_2 \rangle $, and $SNR
\approx \frac{}{\sqrt{3}}\frac{G}{\langle I_1  I_2 \rangle }$.

Of course, after averaging over $N$ independent measurements $SNR(N)= \sqrt{N} SNR $, and if collecting a large amount of samples does not represent the problem, any ghost image/diffraction can be retrieved after a sufficient number of data collections. Hence, if the goal is retrieving information about a macroscopic, stable object, the thermal source represents by far a much better deal than the entangled two-photon source.
Needless to say, if the goal is performing a high sensitivity measurement, or using the ghost imaging scheme as a cryptographic scheme where information need to be hidden to a third party, then the issue of SNR becomes crucial, and the two-photon entangled source may turn out to be the only proper choice.

\section{Spatial coherence properties of the speckle beams}
\label{sec:Spat-coher-prop}
Relevant for the ghost image and the ghost diffraction schemes are the spatial coherence properties of the speckle beams in the object near-field plane, and in the far-field plane with respect to the object. These can be investigated by measuring the autocorrelation function of the intensities. 
In any plane it holds a Siegert-like  factorization formula for thermal statistics \cite{mandel:1995,goodman:2000}: 
\begin{equation}
\langle : I (\xv) I  (\xp) : \rangle=
\langle  I (\xv)\rangle \langle I  (\xp)\rangle +
%+ |r|^4  
%\propto
\frac{1}{M} \left| \Gamma ( \xv , \xp)
\right|^2 \, ,
\label{Siegert}
\end{equation}
where $M$ is the degeneracy factor accounting for the number of
temporal and spatial modes detected. Hence, the properties of the
field correlation function $\Gamma$ can be inferred from the
measurement of the intensity correlation. In particular we will be
interested in the field correlation function at the object near-field
plane $\Gamma_n (\xv, \xp)$, and in the same function at the far-field
plane $\Gamma_f (\xv,\xp)$.

We notice the following equalities, which are a trivial consequence of  the BS input-output relations (\ref{BS})
\begin{eqnarray}
\langle: I_1(\xv) I_1(\xp):\rangle = |t|^4 \langle : I(\xv)
I(\xp):\rangle 
\nonumber\\
= \frac{|t|^2 }{|r|^2 }  \langle I_1(\xv)
I_2(\xp)\rangle  \, \label{I1I1} \\ 
\langle: I_2(\xv) I_2(\xp):\rangle  = |r|^4 \langle : I(\xv)
I(\xp):\rangle 
\nonumber\\
=\frac{|r|^2 }{|t|^2 }   \langle I_1(\xv )
I_2(\xp)\rangle  \, \label{I2I2} \,  
\end{eqnarray} 
where $: \; :$ indicates normal ordering and $I(\xv)$ is the intensity distribution of the speckle beam in the absence of the BS. A part from numerical factors, and from the shot noise contribution at $\xv =\xp $,  in a given plane  the auto-correlation function of each of the two beams coincides with the intensity cross correlation of the 
two beams. 

%%%NEAR FIELD FOURTH ORDER CORRELATION 
\begin{figure}[ht] 
\centerline{    
    \scalebox{.4}{\includegraphics*{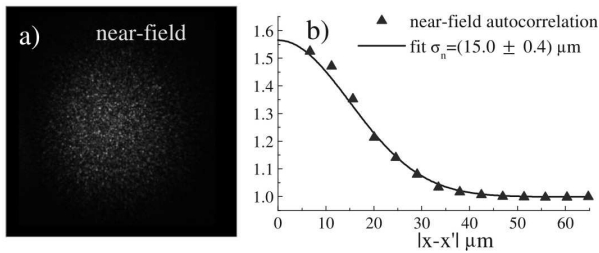}} }
\caption{(a)Instantaneous  intensity distribution $I_2$ of the speckle beam in the near-field plane; (b) Auto-correlation function of the intensity in (a). The full line is a  Gaussian fit of the correlation peak, and the data have been normalized to the baseline values.}
\label{fig:near}
\end{figure}  
%%%%%%
Figure \ref{fig:near}a shows the instantaneous intensity distribution of the reference beam in the setup of Fig.\ref{fig:setup} with the lens F' inserted, so that the data recorded on the CCD are the (demagnified) image of the intensity distribution in the near-field plane.  A large number of speckles 
are visible with a high contrast, due to the short measurement time. According to Van-Cittert Zernike theorem the size of the speckle here is determined by the inverse of the source size (the laser diameter $D_0$) and and by  the distance $z$ from the source \cite{goodman:1975}, $\Delta x_{\rm n}
\propto \lambda z/D_0= 32 \, \mu$m. 
%As we checked, the diaphragm  being close enough to the object plane, its diameter $D$ does not affect much $\Delta x_{\rm n}$. 
Frame (b) in this figure is the radial autocorrelation function (\ref{I2I2}), calculated as a function of the distance $|\xv-\xp|$, normalized to the product of the mean intensities. The baseline corresponds to the product of the mean intensities while the narrow peak located around $|\xv-\xp|=0$ is proportional  to $|\Gamma_n (\xv, \xp)|^2$, where $\Gamma_n$ is the second order {\em field} correlation function at the near-field plane. This peak reflects the spatial coherence properties of the beams at the object plane. Its width is the { \em near-field coherence length} $\Delta x_{\rm n}$ and 
gives an estimate of the speckle size in this plane $\Delta x_{\rm n} \approx 2 m \sigma = 36 \mu$m.  Notice that the peak value is slightly smaller than twice  the baseline 
value, giving a degeneracy factor $M=1.7$.

\begin{figure}[ht] 
\centerline{    
    \scalebox{.4}{\includegraphics*{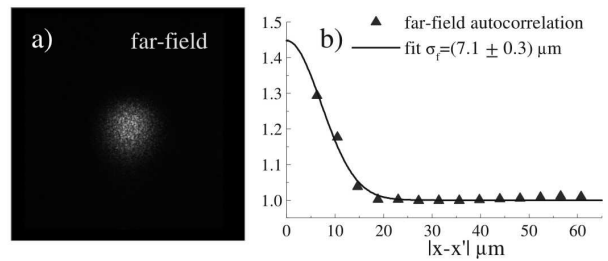}} }
\caption{(a) Instantaneous intensity distribution of the speckles $I_1$ in the far-field plane (b) Auto-correlation function of the intensity in (a). The full line is a  Gaussian fit of the correlation peak, and the data have been normalized to the baseline values.}
\label{fig:far}
\end{figure}

Figure \ref{fig:far} shows the instantaneous intensity distribution (a) and the intensity auto-correlation function (b) in the far-field plane, measured in the focal plane of the lens F. The narrow peak in (b) located around $|\xv-\xp|=0$ is now proportional  to $|\Gamma_f (\xv, \xp)|^2$, and its width  
(the {\em far-field coherence length}) gives an estimate of the speckle size in this plane.
%%%FAR FIELD FOURTH ORDER CORRELATION 
High-contrast speckles are visible also in the far-field plane. The Van-Cittert Zernike theorem can be again invoked to estimate their expected size, being now the source size represented by the diaphragm diameter $D$,
$\Delta x_{\rm f} \propto \lambda F/D \approx 14 \, \mu$m~\cite{goodman:1975}.
This is in good agreement with the estimation from the the correlation function, that gives $\Delta x_{\rm f} = 2 \sigma_f = 14.2 \mu$m.
In this case the peak value of the correlation function in frame (b) gives a degeneracy factor $M=2.2$. This is slightly higher than in the near field because $\Delta x_{\rm f}$ (the size of the spatial mode) is smaller and comparable with the pixel side ($ 6.7  \mu $m.)
\par
Because of the identities in Eqs. (\ref{I1I1}),(\ref{I2I2}), the cross-correlation $\langle I_1 I_2 \rangle$ in the near and in the far-field coincides with the auto-correlation plotted in Fig.\ref{fig:near}b and in Fig.\ref{fig:far}b, respectively. Hence a high degree of mutual spatial correlation is present in both planes, as a consequence  of the spatial incoherence of the light produced by our source. The more incoherent is the light (the smaller the speckles with respect to the beam size), the more localized is the spatial mutual correlation function. The more coherent is the source (the larger the speckles with respect to the beam size), the flatter is the spatial mutual correlation function. As it will become clear in the next two sections, for highly spatially incoherent light, both the ghost diffraction and the ghost image can be retrieved with high resolution. Conversely, in the limit of spatially coherent light no spatial information about the object can be extracted in a {ghost imaging scheme}, that is from the intensity cross-correlation of the two beams as a function of the pixel position in the reference beam.
\par
Summarizing, two aspects of our experiment are crucial i) the spatial incoherence of light, and ii) a measurement time  $ \ll\tau_{coh}$.  Notice that the presence in the near-field of a large number of small speckles inside a broad beam, implies that the light is incoherent also in the far field, because $\Delta x_{\rm f} \propto 1/D$, while the far-field diameter of the beam $ \propto 1/\Delta x_{\rm n}$. 

\section{The ghost diffraction experiment: complementarity between coherence and correlation }
\label{sec:diff}
In this section we focus on the ghost diffraction setup
(Fig.\ref{fig:setup} without the lens $F'$). The object is a double
slit, consisting of a thin needle of 160 $\mu$m diameter inside a
rectangular aperture 690 $\mu$m wide. 

\begin{figure}[ht] 
\centerline{    
    \scalebox{.4}{\includegraphics*{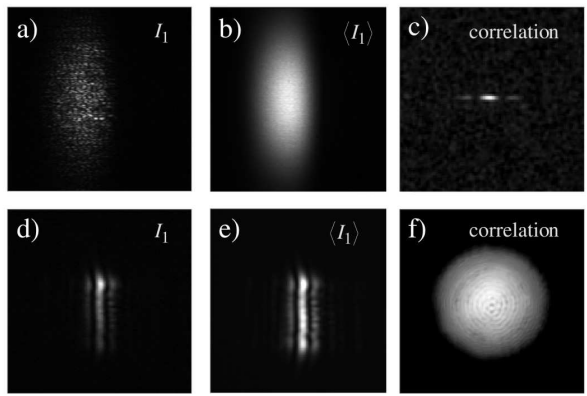}} }
\caption{Ghost diffraction setup: transition from incoherent light to partially coherent light. In the three upper frames (a--c) the source size is $D_0=10 mm$, with near-field speckles $\Delta x_{\rm n} =36 \mu$m. In the three lower frames (d--f) the source size is $D_0=0.1 mm$, with $\Delta x_{\rm n} = 3.2 $mm. (a)and (d):  Instantaneous intensity distribution $I_1$ of the object beam. (b)and (e): Intensity distribution $\langle I_1\rangle $, averaged over 350 shots (c)and (f): Correlation $G(\xv_1, \xv_2)$ as a function of $\xv_2$, for a fixed $\xv_1$, averaged  over 20000 shots.}
\label{fig:duality}
\end{figure}

\begin{figure}[h] 
\centerline{    
    \scalebox{.4}{\includegraphics*{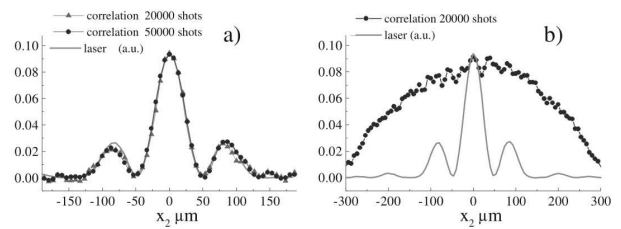}} }
\caption{Horizontal sections of the correlation $G(\xv_1, \xv_2)$ as a function of $x_2$, for a fixed $\xv_1$ (see Fig.\ref{fig:duality}c, f).  (a) Is  the case of incoherent light, $D_0=10 $mm; the data  are obtained with an average over 20000 shots (triangles) and  50000 shots (circles). (b) Is the case  of partially coherent illumination,$D_0=0.1$mm (20000 shots). The light full line is for comparison the diffraction pattern observed with a laser.}
\label{fig:secdiff}
\end{figure}

In a first set of measurements the source size is $D_0 =10 mm$, and the object is illuminated by a large number of speckles whose size $\Delta x_{\rm n} =36 \mu$m is much smaller than the slit separation. The light is {\em spatially incoherent} as described in the previous section. The results are shown in the first row of Fig.\ref{fig:duality}. Frame (a) is the instantaneous intensity distribution of the object beam, showing a speckled pattern, with no interference fringes from the double slit, as expected for incoherent illumination \cite{mandel:1995}. 
%%%DUALITY FIGURE 
%%%%%%%%%%%%
At a closer inspection, the shape of the speckles resembles the interference pattern of the double slit, but since these speckles move randomly in the transverse plane from shot to shot, an average over several shots displays a homogeneous broad spot (Fig \ref{fig:duality}(b)). Frame (c) is a plot of $G(\xv_1, \xv_2)$ as a function of the reference pixel position $\xv_2$, and shows the result of correlating the intensity distribution in the reference arm with the intensity collected from a single fixed pixel in the object arm. Notice that at difference to what was done in \cite{ferriexp-osa}, no spatial average \cite{bache:2004a} is here employed: this makes the convergence rate slower but the scheme is closer to the spirit of ghost diffraction in which the information is retrieved by only scanning the reference pixel position. The {\em ghost diffraction} pattern emerges after a few thousands of averages, and is well visible after 20000 averages. This is confirmed  by the data of  Fig.\ref{fig:secdiff}a which compare the horizontal section of the diffraction pattern from a correlation measurement to that obtained with laser illumination. 
%%%CORRELATION  SECTIONs G
%%%%%%%%%%%%
%%%INTENSITY  SECTIONS I
\begin{figure}[h] 
\centerline{    
    \scalebox{.4}{\includegraphics*{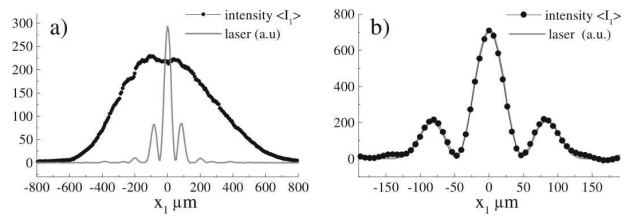}} }
\caption{Horizontal sections of the average intensity distribution  $\langle I_1 (\xv_1) \rangle$ in the object arm (see Fig.\ref{fig:duality}b, e).  . (a) Is obtained  for incoherent light,with $D_0=10 $mm (350 shots), while  (b) plots the case of  partially coherent illumination, with $D_0=0.1$mm (500 shots).The light full line is for comparison the diffraction pattern observed with a laser.}
\label{fig:secint}
\end{figure}  
%%%%%%%%%%%%
In a second set of measurements the source size is reduced to $D_0 =0.1 mm$, by inserting a  small pinhole after the ground glass. As a result, the spatial coherence of the light illuminating the object is increased. As the speckle size at the diaphragm $D$ is now $\approx 3 mm$ , on average the object is illuminated by a single speckle of size much larger than the slit separation. 
The results are reported in the second row of Fig.\ref{fig:duality}.
As expected \cite{mandel:1995} the interference fringes are now visible in the instantaneous intensity distribution of the object beam 1 [frame (d)], and become sharper after averaging over some hundreds of shots [frame (e)]. Notice that the shape of the interference pattern is now elongated in the vertical direction, because the light emerging from the small source is not collimated.   Horizontal sections of $\langle I_1 \rangle$ , plotted in Fig.\ref{fig:secint}b, show a very good agreement with the diffraction pattern from laser illumination. Instead, no interference fringes at all appear in the correlation function of the intensities in the two arms, when plotted as a function of $\xv_2$ [frame (f)]. 
Notice that in this set of measurements  the turbid medium was removed
in order to increase the power. This is feasible in this case, because
the very small size of the source allows a large number of independent
patterns to be generated in a single tour of the glass disk. 

Figures \ref{fig:duality},\ref{fig:secdiff},\ref{fig:secint} evidence
a remarkable complementarity between the observation of interference
fringes in the correlation function (ghost diffraction), and in the
intensity distribution of the object beam (ordinary diffraction). It
also shows the fundamental role played by the spatial incoherence of
the source in producing a ghost diffraction pattern: the more
incoherent is the source, the more the two beams are spatially
correlated and the more information about the object is available in
the ghost diffraction pattern. The more coherent is the source, the
flatter is the spatial correlation function of the two beams and the
less information about the object is contained  in the ghost
diffraction. This is completely analogous to the complementarity
between the one-photon and two-photon interference in Young's double
slit experiments with photons from a PDC source
\cite{abouraddy:2001c}, which was explained as a complementarity
between coherence and entanglement. In our case of thermal beams, the
complementarity is rather between coherence and spatial correlation,
showing that also in this respect the classical spatial correlation
produced by splitting thermal light play the same role as entanglement
of PDC photons.  

These results can be easily understood by using the formalism
developed in Sec.\ref{sec:formal}, and in particular by inspection of
Eq.(\ref{eq12})for the correlation function of the intensity
fluctuations $G(\xv_1, \xv_2)$. In the limit of spatially coherent
light the field correlation function $\Gamma_n (\xv_1, \xv_2)$ becomes
constant in space in the region of interest, and the two integrals in
Eq.(\ref{eq12}) factorize to the product of two ordinary imaging
schemes, showing the diffraction pattern of the object only in the
object arm 1. As a result, by plotting the correlation as a function
of $\xv_2$, no object diffraction pattern can be observed, that is, no
ghost diffraction occurs. The same observation can be made with
respect to $\Gamma_{\rm pdc} (\xv_1, \xv_2)$, and $G_{\rm pdc}(\xv_1,
\xv_2)$, explaining thus the analogy between the role of light
coherence in the PDC and in the thermal case. 

In general, the result of a correlation measurement is obtained by inserting into Eq. (\ref{eq12}) the propagators
that describe the ghost diffraction setup: $h_1(\xv_1,\xp_1)=(\im\lambda F)^{-1} e^{-\frac{2\pi\im}{\lambda F}
  \xv_1 \cdot \xp_1}    T(\xp_1)$, with 
$T(\xv)$ being the object transmission function, and 
$h_2(\xv_2,\xp_2)=(\im\lambda F)^{-1} e^{-\frac{2\pi\im}{\lambda F}
  \xv_2 \cdot \xp_2} $. We get   
\begin{equation}
G (\xv_1, \xv_2) \propto
%\frac{RT}{ (\lambda f)^2}
\left|
\int {\rm d} \vec{\xi}
 \,
\tilde{T} \left[ (\xv_1-\vec{\xi}) \frac{2\pi}{\lambda F} \right] \Gamma_f (\xv_2, \vec{\xi} ) 
\right|^2 \, ,
\label{diff1} 
\end{equation}
where $\tilde T (\q) = \int \frac{{\rm d} \xv}{2\pi} e^{-\im \q \cdot
  \xv} T(\xv)$ is the amplitude of the diffraction pattern from the
object. The result of a correlation measurement is a convolution of the
diffraction pattern amplitude with the second order correlation function in the far-field. Hence the {\em far-field coherence length} determines the spatial resolution in the ghost diffraction scheme: the smaller the far-field speckles, the better resolved is the pattern. In the limit of speckles much smaller than the scale of variation of the diffraction pattern
\begin{equation}
G (\xv_1, \xv_2) \to 
\left|  \tilde{T} \left[ (\xv_1-\xv_2) \frac{2\pi}{\lambda F} \right] 
\right|^2 
\left|
\int {\rm d} \vec{\xi}
 \Gamma_f( \vec{\xi},\xv_2) 
\right|^2\; ,
\label{diff22}
\end{equation}

Since the amplitude of the object diffraction pattern is involved in
Eq.(\ref{diff1}), ghost diffraction of a pure phase object can be
realized with spatially incoherent pseudo-thermal beams, a possibility which was questioned in a recent letter \cite{abouraddy:2004}.

\section{The ghost image: tradeoff between resolution and visibility} \label{sec:image}
\begin{figure}[ht] 
\centerline{    
    \scalebox{.4}{\includegraphics*{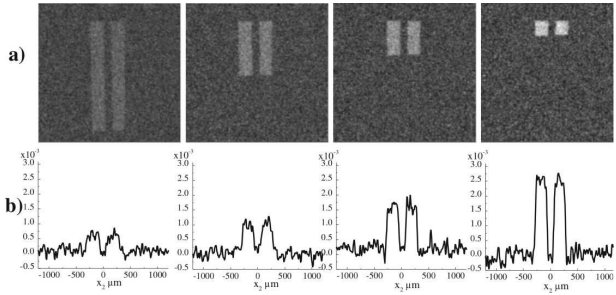}} }
\caption{Reconstruction of the object image via correlation measurements  (Fig.\ref{fig:setup}, with the lens $F^{\prime}$ inserted). (a)Cross-correlation of the intensity distribution of the reference arm with the total photo counts in the object arm, as a function of $\xv_2$. (statistics over 10000 CCD frames). In the sequence of frames the object area is progressively reduced, and correspondingly a enhancement of the visibility can be observed (b) Horizontal sections of the images in (a), with the correlation normalized to $\langle I_1 I_2 \rangle$, so that the vertical scales gives the visibility. }
\label{fig:images}
\end{figure}

By simply inserting the lens $F'$ in the reference arm (see Fig.1), without changing anything in the object arm, we  now pass to the ghost image. As it was predicted in \cite{thermal-oe}, and experimentally demonstrated in \cite{ferriexp-osa}, the result of cross-correlating the intensities of the reference and object arm is now the image of the object, shown e.g. in Fig. \ref{fig:images}a.
\par
Two issues are important in any imaging scheme: the spatial resolution
and the signal-to-noise ratio. 

As pointed out in \cite{thermal-oe,ferriexp-osa}, the resolution capabilities of the ghost image setup are determined by the near field coherence length $\Delta x_{\rm n}$ (the size of the near-field speckles). This can be easily understood by inserting the  
propagator $h_2 (\xv_2,\xp_2) = m\delta
(m \xv_2+\xp_2) $, into Eq. (\ref{eq12}):
\begin{equation}
G (\xv_1, \xv_2) \propto 
%\frac{RT}{ (\lambda f)^2}
\left|
\int {\rm d} \xv_1'
 \Gamma_n (\xv_1', -m \xv_2) 
T^*( \xv_1') e^{\im \frac{2\pi}{\lambda f}\xv_1 \cdot \xv_1'}
\right|^2 \, ,
\label{image1} 
\end{equation}
which shows that the result of a correlation measurement in this setup is a convolution of the object transmission function with the near-field correlation function $\Gamma_n$. In the following we shall consider a {\em bucket } detection scheme, where the reference beam intensity $I_2$ is correlated to the total photon counts in the object arm, that is, in practice to the sum   of photo counts over a proper set of pixels. This makes the imaging incoherent \cite{bache:2004a}, because it amounts to measuring 
\begin{equation}
\int {\rm d} \xv_1 G (\xv_1, \xv_2) =
\int {\rm d} \xv_1' \left|
 \Gamma_n (\xv_1', -m \xv_2) \right|^2 
\left| T( \xv_1') \right|^2 \, .
\label{imagebucket}
\end{equation}
If we take the limit of spatially coherent light, where $\Gamma_n (\xv_1', -m \xv_2)$ can be considered as constant over the whole beam size,  
Both Eqs. (\ref{image1}),(\ref{imagebucket}) show that no information about the object image can be obtained by scanning $\xv_2$. Conversely in the limit of spatially incoherent light where the speckle size is much smaller than the scale of variation of the object, $\Gamma_n (\xv_1', -m \xv_2) \approx \delta(\xv_1'+ m \xv_2) $,
and both Eqs. (\ref{image1}),(\ref{imagebucket})  converge to $\left|  T \left( -m \xv_2  \right) \right|^2 \times {\rm const.} $.

Concerning the signal-to-noise ratio, the discussion in Sec. \ref{sec:formal} pointed out that it is determined by the image visibility. We have studied the visibility of the ghost image of a double slit in a sequence of measurements where the vertical size of the apertures was progressively reduced, while leaving unchanged their horizontal size and separation. This is shown in Fig.\ref{fig:images}a, where all the frames display the correlation function  measured in a bucket detection scheme as a function of the reference pixel position $\xv_2$. Despite the fact that all the frames have  been  obtained with the same number of averages $N=10000$, the sequence display a remarkable enhancement of the visibility as the object area is reduced. 
%%%IMAGE SEQUENCE
%%%%%
This enhancement is more clearly visible in the horizontal sections
plotted in Fig. \ref{fig:images}b, where in each point the correlation
function has been normalized to $\langle I_1 I_2\rangle$, so that the
numbers on the vertical axis give directly the visibility [see Eq.
(\ref{visibility})]. We notice that the visibility increases as the
object area decreases, and correspondingly the the
signal-to-noise-ratio increases, as expected.
\par
\begin{figure}[h] 
\centerline{    
    \scalebox{.3}{\includegraphics*{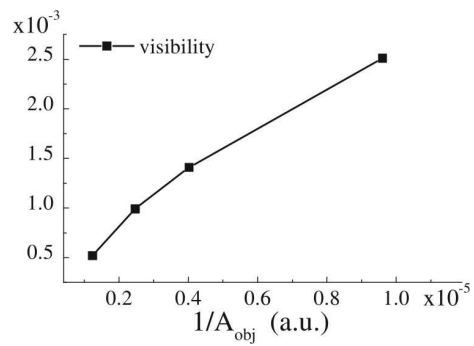}} }
\caption{Visibility of the ghost image as a function of the inverse of the transmissive area of the object, showing an increase of the visibility by reducing the object area.}
\label{fig:visibility}
\end{figure}
\begin{figure}[h] 
\centerline{    
    \scalebox{.2}{\includegraphics*{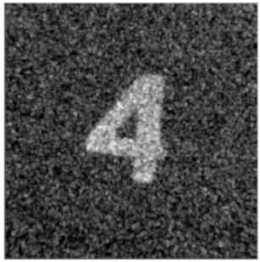}} }
\caption{Ghost image of the number 4, retrieved from the correlation function after averaging over 30000 shots.}
\label{fig:quattro}
\end{figure}

In order to understand the origin of the behaviour shown in Fig.\ref{fig:images}, we need first to consider  Eq.(\ref{imagebucket}), that gives the correlation function in the bucket detection scheme. Let us assume that the object simply transmits the light over a region of area $S_{obj}$ and stops it elsewhere. By assuming that the coherence length $\Delta x_{\rm n}$ is smaller than the object features, as it is necessary for the object to be resolved, the integrand at r.h.s of (\ref{imagebucket}) can be non-zero for $\xp_1$ in a region located around  $\xv_2$, of area $A_{coh}$, where $A_{coh}$ is the {\em coherence area} $\propto \Delta x_{\rm n}^2$. Thus the correlation scales as the coherence area. 
\begin{equation}
\int {\rm d} \xv_1 G(\xv_1,\xv_2) \propto A_{coh}
\end{equation}
Conversely, it is not difficult to see that in the bucket detection scheme
\begin{equation}
 \int {\rm d } \xv_1 \langle I_1(\xv_1) \rangle = \int {\rm d} \xp_1 \left |T (\xp_1)\right|^2 \langle 
I_n (\xp_1) \rangle \propto A_{obj} \, , 
\end{equation}
 where $ \langle I_n (\xp_1)\rangle $ is the average intensity distribution of the light illuminating the object, that can be taken as roughly uniform on the object area (the speckles average to a broad uniform light spot, as shown in Fig. \ref{fig:duality}b). Hence the ratio of the correlation to the background scales as:
\begin{equation}
\frac{ \int {\rm d } \xv_1 G(\xv_1,\xv_2) }{ \int {\rm d } \xv_1 \langle I_1(\xv_1)\rangle \langle  I_2(\xv_2) \rangle  } \propto \frac{A_{coh}}{A_{obj}} \; .
\label{ratio}
\end{equation}
This formula is reminiscent of role of the mode degeneracy in Eq.(\ref{Siegert}), and indeed it reflects the fact that in a bucket detection scheme the number of spatial modes detected is proportional to $A_{obj}/A_{coh}$, which represents a degeneracy factor that reduces the visibility of the correlation with respect to the background.
The ratio in Eq. (\ref{ratio}) is usually small, so that the visibility of the ghost image roughly coincides with it. Thus the visibility roughly scales as the ratio of the coherence area to the object transmissive area: the larger are the object dimensions with respect to the speckles, that is, the more incoherent is the light illuminating the object, the worse is the visibility of the ghost image. This is confirmed by the plot in Fig. \ref{fig:visibility}, showing how the visibility increases with the inverse of the object area.
%%%VISIBILITY PLOT 
%%%%%
This rather counter-intuitive result also implies that a better
resolution can be achieved only at the expenses of the visibility,
since the resolution is determined by the speckle size. Hence, complex
images that need small speckles to be resolved in their details have a
lower signal-to-noise ratio than simple images which can be resolved
with relatively large speckles (see also \cite{valencia:2004} for a
similar conclusion). This, however, does not prevent from retrieving
more complex images (see e.g. Fig.~\ref{fig:quattro}), provided that a
larger number of data acquisitions are performed.
%%%QUATTRO 
%%%%%

%%%%%%%%%%%%%%%%%%%%%%%
\section{Numerical results}
\label{sec:Numerical-results}
In this section we use a numerical model for simulating the speckle
pattern created by the ground glass and the turbid medium, as to
support the results of the previous section. The thermal field is
created by generating a noisy field with huge phase fluctuations. We
then multiply the noisy field with a Gaussian profile and a subsequent
Fourier transformation gives what corresponds to the near field; the
width of the Gaussian then controls the near-field speckle size. The
far-field speckle size is controlled independently as in the
experiment: after generating the near field, a diaphragm of diameter
$D$ is introduced, beyond which only vacuum fluctuations appear; $D$
then controls the far-field speckle size. The speckle field
transmitted by the diaphragm is then impinged on a 50/50 BS, with
vacuum fluctuations entering the unused port, giving the two desired
correlated beams. We neglect the temporal statistics, since we assume
that the short measurement time of the experiment provides a speckle
pattern frozen in time. We should finally mention that a Wigner
formalism is used to describe the quantized fields, as described in
detail in \cite{brambilla:2004}.

\begin{figure}[ht]
  \begin{center}
    \includegraphics[width=8.5cm]{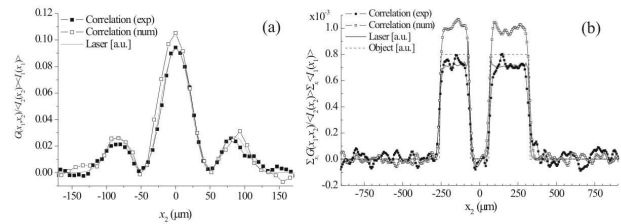}
  \caption{Comparing a two dimensional numerical simulation of the
    experiment, by showing the
    correlation of intensity fluctuations normalized to the product of
    the beam intensities. (a) The ghost diffraction case
    $G(\vec{x}_1,\vec{x}_2)/\langle I_1(\vec{x}_1)\rangle\langle
    I_2(\vec{x}_2)\rangle$. (b) The ghost image case $\int {\rm d} x_1
    G(x_1,x_2)/ \langle I_2(x_2)\rangle\int {\rm d} x_1\langle
    I_1(x_1)\rangle$. In both cases the numerics and experiment are
    real units, and are as reference compared to the data obtained by
    coherent laser illumination of the object. The averages are done
    over $2\cdot 10^4$
    realizations. In the numerics $\Delta x_{\rm n}=34~\mu$m and
    $\Delta x_{\rm f}=12~\mu$m.}
   \label{Fig:ff-nf-num-exp}
  \end{center}
\end{figure}

Initially, let us briefly show that the numerical simulations are able
to describe very precisely the experiment. In
Fig.~\ref{Fig:ff-nf-num-exp} is shown the results of two-dimensional
(2D) simulations with all parameters kept as close as possible to the
experiment. This includes near-field and far-field speckle sizes,
object and aperture sizes, as well as number of realizations. Both the
ghost diffraction pattern [Fig.~\ref{Fig:ff-nf-num-exp}(a)] and the
ghost image [Fig.~\ref{Fig:ff-nf-num-exp}(b)] show very good agreement
with the experimental recorded data (using the small-speckle setup of
Secs.~\ref{sec:Spat-coher-prop}-~\ref{sec:image}). We stress that this
comparison is not in arbitrary units.

\begin{figure}[t]
  \begin{center}
    \includegraphics[width=6.6cm]{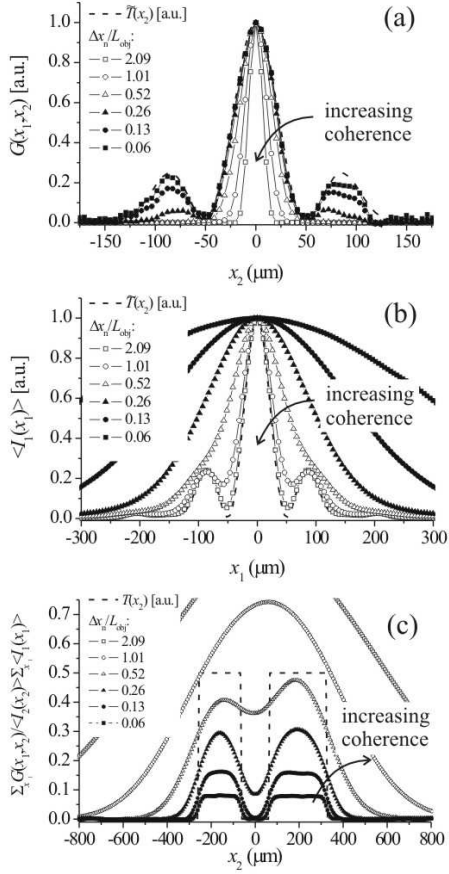}
  \caption{1D numerical simulation of the experiment
    showing the transition from incoherent to coherent illumination of
    the object, realized by changing the near-field speckle size
    $\Delta x_{\rm n}$. (a) shows the normalized correlation of
    intensity fluctuations in the ghost diffraction case, while (b)
    shows the normalized $\langle I_1(x_1)\rangle$ as observed
    directly in the object arm. (c) shows the correlation of intensity
    fluctuations in the ghost image case, normalized to the beam
    intensities $\int {\rm d} x_1 G(x_1,x_2)/ \langle
    I_2(x_2)\rangle\int {\rm d} x_1\langle I_1(x_1)\rangle$. The
    averages are done over $10^5$ realizations. The object mimics the
    experimental one, implying $L_{\rm obj}=530~\mu$m. $\Delta x_{\rm
    f}=12~\mu$m. }
   \label{Fig:trans-num}
  \end{center}
\end{figure}

In Sec.~\ref{sec:diff} we showed experimentally the behaviour of the
system when using either coherent or incoherent beams to investigate
the diffraction properties of an object. In order to investigate
better the actual transition from coherent to incoherent illumination
of the object, we have carried out numerical simulations that are
presented in Fig.~\ref{Fig:trans-num}.  We have there kept $D=3$~mm
and then for each simulation changed $\Delta x_{\rm n}$. The
simulation only includes one spatial direction (1D), and therefore the
coherence properties of the beam is governed by the ratio $\Delta
x_{\rm n}/L_{\rm obj}$, where $L_{\rm obj}$ is the 1D equivalent of
$A_{\rm obj}$. Thus, the smaller $\Delta x_{\rm n}/L_{\rm obj}$ the
more incoherent is the beam impinging on the object. For small
speckles ($\Delta x_{\rm n}/L_{\rm obj}\ll 1$) the beams are spatially
incoherent, implying a strong spatial correlations between the beams:
the ghost diffraction is observed in the correlation
[Fig.~\ref{Fig:trans-num}(a)]. In contrast, no diffraction pattern can
be observed directly in the object beam [Fig.~\ref{Fig:trans-num}(b)].
As $\Delta x_{\rm n}$ is increased by generating bigger speckles the
beams become more and more spatially coherent ($\Delta x_{\rm
  n}/L_{\rm obj}\simeq 1$): the ghost diffraction disappears gradually
[Fig.~\ref{Fig:trans-num}(a)], while the diffraction pattern starts to
appear from the direct observation of the object beam
[Fig.~\ref{Fig:trans-num}(b)].  Figure~\ref{Fig:trans-num}(c) shows
what happens for the ghost image during this transition: the
incoherence for small $\Delta x_{\rm n}$ implies that a ghost image of
the object can be observed, and it disappears gradually while
increasing the coherence.  

\begin{figure}[h]
  \begin{center}
    \includegraphics[width=8.5cm]{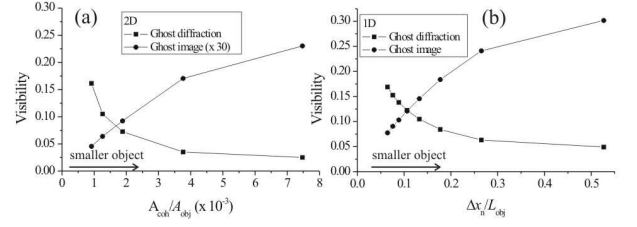}
  \caption{Numerical simulations of the experiment
    showing how the object size effects the visibility $V$. We kept
    the speckle sizes constant but varied the width of the two slits.
    (a) is the 2D case showing $V$ as function of the speckle area
    relative to the object area. Note that the ghost image visibility
    has been multiplied by a factor of 30, and that the object length
    perpendicular to the slits was kept constant. (b) is the
    1D case, showing $V$ as function of the speckle size relative to
    the object length.  $\Delta x_{\rm n}=34~\mu$m and $\Delta x_{\rm
      f}=12~\mu$m.}
   \label{Fig:vis-num}
  \end{center}
\end{figure}

We saw in Sec.~\ref{sec:image} that the visibility of the ghost image
became better as the object area was reduced, cf.
Fig.~\ref{fig:visibility}. To investigate this phenomenon in general
we show in Fig.~\ref{Fig:vis-num} how the object size\footnote{Note
  that we here keep the length perpendicular to the slits constant but
vary the width of the slits. Experimentally, this would correspond to
maintaining the needle but changing the width of the surrounding
aperture.} effects the
visibility of the information. The trend we saw in the experiment is
repeated in the numerics: in Fig.~\ref{Fig:vis-num}(a) the ghost image
visibility increases as the object size decreases because fewer modes
are transmitted. In Fig.~\ref{Fig:vis-num}(b) the simulation is
repeated for the 1D case with a similar result. However, since in the
1D case much fewer modes are transmitted by the object the visibility
is much higher. We have also in Fig.~\ref{Fig:vis-num} plotted the
visibility of the ghost diffraction fringes, and we observe that -- in
contrast to the ghost image case -- the visibility decreases as the
object size is decreased. %(a result reported also in \cite{cai:2004d}).
This is is to be expected because for the diffraction pattern the
modes transmitted by the object interfere coherently so
$G(\vec{x}_1,\vec{x}_2)\propto A_{\rm obj}^2$ (for the 2D case). In
contrast, for the mean intensity the modes interfere incoherently so
$\langle I_1(\vec{x}_1)\rangle\propto A_{\rm obj}$. Thus
$G(\vec{x}_1,\vec{x}_2)/\langle I_1(\vec{x}_1)\rangle\langle
I_2(\vec{x}_2)\rangle\propto A_{\rm obj}$: the bigger the object the
better the visibility of the information. We also note that there is
basically no difference between the 1D and 2D results for the ghost
diffraction visibility. This is because a similar argument can be done
for the 1D case showing $G(x_1,x_2)/\langle I_1(x_1)\rangle\langle
I_2(x_2)\rangle\propto L_{\rm obj}$. Finally, we have checked that if
the far-field speckle size $\Delta x_{\rm f}$ is varied and all other
parameters are kept fixed, then the visibility of the diffraction
fringes increases as $\Delta x_{\rm f}$ is increased: again a tradeoff
between resolution and visibility is found.

\section{Conclusions} 
\label{sec:Conclusions}
The experimental results reported in this paper confirms that correlated imaging can be performed with a
classical thermal source. A remarkable complementarity between spatial coherence and correlation is predicted and confirmed by experiments and numerical simulations. By changing the coherence of the
speckle field at the object plane from incoherent to coherent
(measured relative to the object dimensions), the object diffraction
pattern reconstructed from correlations disappears while it appears in
the far field intensity distribution measured in the object arm.  We
also discussed from a quantitative point of view the issue of the
visibility of the correlated imaging scheme.  We showed that the
visibility of the object image is proportional to the ratio between
the object area and the field coherence area at the object plane.
This means that a tradeoff between resolution and visibility exists: a
better visibility can be obtained only at the expense of a lower
resolution and vice versa. However, the experiment clearly shows that
a fairly good resolution can be achieved since the problem of low
visibility can be overcome by performing a sufficiently large number
of averages. 
  
\section*{Acknowledgments} This work was carried out in the framework of the 
FET project QUANTIM of the EU, of the PRIN project of MIUR "Theoretical
study of novel devices based on quantum entanglement", and of the
INTAS project "Non-classical light in quantum imaging and continuous
variable quantum channels". M.B. acknowledges financial support from
the Carlsberg Foundation.
%%%%%%%%%%%%%%%%%%%%%
%\input{submit.bbl}
%
\bibliographystyle{C:/texmf/tex/latex/revtex/osa}
\bibliography{d:/Projects/Bibtex/literature}

\begin{thebibliography}{10}

\bibitem{strekalov:1995}
D.~V. Strekalov, A.~V. Sergienko, D.~N. Klyshko, and Y.~H. Shih, ``Observation
  of Two-Photon "Ghost" Interference and Diffraction,'' Phys. Rev. Lett. {\bf
  74,} 3600--3603 (1995).

\bibitem{ribeiro:1994}
P.~H. {Souto Ribeiro}, S. Padua, J.~C. {Machado da Silva}, and G.~A. Barbosa,
  ``Controlling the degree of visibility of Young's fringes with photon
  coincidence measurements,'' Phys. Rev. A {\bf 49,} 4176--4179 (1994).

\bibitem{pittman:1995}
T.~B. Pittman, Y.~H. Shih, D.~V. Strekalov, and A.~V. Sergienko, ``Optical
  imaging by means of two-photon quantum entanglement,'' Phys. Rev. A {\bf 52,}
  R3429--R3432 (1995).

\bibitem{abouraddy:prl-josab-osa}
A. F. Abouraddy, B. E. A. Saleh, A. V. Sergienko, and M. C. Teich, "Role of
  Entanglement in two photon imaging," Phys. Rev. Lett. {\bf 87}, 123602
  (2001); "Entangled-photon Fourier optics," J. Opt. Soc. Am. B {\bf 19}, 1174
  (2002).

\bibitem{bennink:2002a}
R.~S. Bennink, S.~J. Bentley, and R.~W. Boyd, ``"Two-Photon" Coincidence
  Imaging with a Classical Source,'' Phys. Rev. Lett. {\bf 89,} 113601 (2002).

\bibitem{gatti:2003}
A. Gatti, E. Brambilla, and L.~A. Lugiato, ``Entangled Imaging and
  Wave-Particle Duality: From the Microscopic to the Macroscopic Realm,'' Phys.
  Rev. Lett. {\bf 90,} 133603 (2003).

\bibitem{bennink:2004}
R.~S. Bennink, S.~J. Bentley, R.~W. Boyd, and J.~C. Howell, ``Quantum and
  Classical Coincidence Imaging,'' Phys. Rev. Lett. {\bf 92,} 033601 (2004).

\bibitem{brambilla:2004}
E. Brambilla, A. Gatti, M. Bache, and L.~A. Lugiato, ``Simultaneous near-field
  and far field spatial quantum correlations in the high gain regime of
  parametric down-conversion,'' Phys. Rev. A {\bf 69,} 023802 (2004).

\bibitem{howell:2004}
J.~C. Howell, R.~S. Bennink, S.~J. Bentley, and R.~W. Boyd, ``Realization of
  the Einstein-Podolsky-Rosen Paradox Using Momentum- and Position-Entangled
  Photons from Spontaneous Parametric Down Conversion,'' Phys. Rev. Lett. {\bf
  92,} 210403 (2004).

\bibitem{dangelo:2004}
M. D'Angelo, Y.-H. Kim, S.~P. Kulik, and Y. Shih, ``Identifying Entanglement
  Using Quantum Ghost Interference and Imaging,'' Phys. Rev. Lett. {\bf 92,}
  233601 (2004), quant-ph/0401007.

\bibitem{thermal-oe}
A. Gatti, E. Brambilla, M. Bache, and L. A. Lugiato, "Ghost imaging with
  thermal light: comparing entanglement and classical correlation," Phys. Rev.
  Lett. {\bf 93}, 093602 (2004), quant-ph/0307187; "Correlated imaging, quantum
  and classical," Phys. Rev. A {\bf 70}, 013802 (2004), quant-ph/0405056.

\bibitem{ferriexp-osa}
D. Magatti, F. Ferri, A. Gatti, E. Brambilla, M. Bache, and L. A. Lugiato,
  "Experimental evidence of high-resolution ghost imaging and ghost diffraction
  with classical thermal light," quant-ph/0408021; F. Ferri, D. Magatti, A.
  Gatti, E. Brambilla, M. Bache, and L. A. Lugiato, "High-resolution ghost
  image and ghost diffraction experiments with thermal light," to appear in
  Phys. Rev. Lett. (2004).

\bibitem{cai:2004d}
Y. Cai and S.-Y. Zhu, ``Ghost interference with partially coherent radiation,''
  Optics Letters {\bf 29,} 2716 (2004).

\bibitem{cao:2004}
D.-Z. Cao, Z. Li, Y.-H. Zhai, and K. Wang, ``One-Photon and Two-Photon
  Double-Slit Interferences in Spontaneous and Stimulated Parametric
  Down-Conversions,''   (2004), quant-ph/0401109.

\bibitem{cheng:2004}
J. Cheng and S. Han, ``Incoherent Coincidence Imaging and Its Applicability in
  X-ray Diffraction,'' Phys. Rev. Lett. {\bf 92,} 093903 (2004).

\bibitem{scarcelli:2004}
G. Scarcelli, A. Valencia, and Y. Shih, ``Experimental study of the momentum
  correlation of a pseudothermal field in the photon-counting regime,'' Phys.
  Rev. A {\bf 70,} 051802(R) (2004), quant-ph/0409210.

\bibitem{valencia:2004}
A. Valencia, G. Scarcelli, M. {D'Angelo}, and Y. Shih, ``Two-photon imaging
  with thermal light,'' Phys. Rev. Lett. {\bf 94,} 063601 (2005),
  quant-ph/0408001.

\bibitem{abouraddy:2004}
A.~F. Abouraddy, P.~R. Stone, A.~V. Sergienko, B.~E.~A. Saleh, and M.~C. Teich,
  ``Entangled-Photon Imaging of a Pure Phase Object,'' Phys. Rev. Lett. {\bf
  93,} 213903 (2004).

\bibitem{boto:2000}
A.~N. Boto, P. Kok, D.~S. Abrams, S.~L. Braunstein, C.~P. Williams, and J.~P.
  Dowling, ``Quantum Interferometric Optical Lithography: Exploiting
  Entanglement to Beat the Diffraction Limit,'' Phys. Rev. Lett. {\bf 85,} 2733
  (2000).

\bibitem{dangelo:2001}
M. {D'Angelo}, M.~V. Chekhova, and Y. Shih, ``Two-Photon Diffraction and
  Quantum Lithography,'' Phys. Rev. Lett. {\bf 87,} 013602 (2001).

\bibitem{wang:2004}
K. Wang and D.-Z. Cao, ``Subwavelength coincidence interference with classical
  thermal light,'' Phys. Rev. A {\bf 70,} 041801R (2004), quant-ph/0404078.

\bibitem{scarcelli:2004a}
G. Scarcelli, A. Valencia, and Y. Shih, ``Two-photon interference with thermal
  light,'' Europhys. Lett. {\bf 68,} 618--624 (2004), quant-ph/0410217.

\bibitem{hanburybrown:1956}
R. Hanbury-Brown and R.~Q. Twiss, ``Correlations between photons in 2 coherent
  beams of light,'' Nature (London) {\bf 177,} 27 (1956).

\bibitem{goodman:1975}
J.~W. Goodman, ``Statistical properties of laser speckle patterns,''  in {\em
  Laser speckle and related phenomena}, Vol.~9 of {\em Topics in Applied
  Physics}, D. Dainty, ed., (Springer, Berlin, 1975), \ p.\ 9.

\bibitem{Martienssen}
W. Martienssen and E. Spiller, ``Coherence and Fluctuations in Light Beams,''
  Am. J. Phys. {\bf 32,} 919 (1964).

\bibitem{Arecchi}
F.~T. Arecchi, ``Measurement of the Statistical Distribution of Gaussian and
  Laser Sources,'' Phys. Rev. Lett. {\bf 15,} 912 (1965).

\bibitem{cheng:2004a}
J. Cheng and S. Han, ``Theoretical investigation of the quantum noise in ghost
  imaging,''   (2004), quant-ph/0408123.

\bibitem{mandel:1995}
L. Mandel and E. Wolf, {\em Optical coherence and quantum optics} (Cambridge,
  New York, 1995).

\bibitem{goodman:2000}
J.~W. Goodman, {\em Statistical optics} (Wiley Classic Library, New York,
  2000).

\bibitem{bache:2004a}
M. Bache, E. Brambilla, A. Gatti, and L.~A. Lugiato, ``Ghost imaging schemes:
  fast and broadband,'' Opt. Express {\bf 12,} 6067 (2004), quant-ph/0409215.

\bibitem{abouraddy:2001c}
A.~F. Abouraddy, M.~B. Nasr, B.~A. Saleh, A.~V. Sergienko, and M.~C. Teich,
  ``Demonstration of the complimentarity of one- and two-photon interference,''
  Phys. Rev. A {\bf 63,} 063803 (2001).

\end{thebibliography}

\end{document}